\documentclass[11pt]{article}
\pdfoutput=1

\usepackage[T1]{fontenc}
\usepackage[utf8]{inputenc}
\usepackage{authblk}

\usepackage{graphicx}
\usepackage{slashed}
\usepackage{epstopdf}
\usepackage[body={17.5cm, 21cm},right=2cm]{geometry}
\usepackage{amssymb}
\usepackage{amsmath}
\usepackage{booktabs}

\numberwithin{equation}{section}

\newcommand\eea{\end{eqnarray}}
\newcommand\bea{\begin{eqnarray}}

\def\beq{\begin{equation}}
\def\eeq{\end{equation}}

\newcommand{\be}{\begin{equation}}
\newcommand{\ee}{\end{equation}}
\newcommand{\ba}{\begin{align}}
\newcommand{\ea}{\end{align}}
\newcommand{\bg}{\begin{gather}}
\newcommand{\eg}{\end{gather}}
\newcommand{\bseq}{\begin{subequations}}
\newcommand{\eseq}{\end{subequations}}

\begin{document}

\begin{titlepage}
\title{\LARGE\bfseries Measuring SUSY quartics from stop decays}
\author{Anson Hook}
\affil{\small\slshape School of Natural Sciences\\
 Institute for Advanced Study\\
Princeton, NJ 08540, USA}
\maketitle
\thispagestyle{empty}

\begin{abstract}

In supersymmetric theories, there is a scalar quartic related to the top and bottom yukawas that gives LHC observable decays of the stop or sbottom.  After electroweak symmetry breaking, this quartic induces the decay of the stop (sbottom) to a charged Higgs and the sbottom (stop) without allowing for the corresponding channel involving the W boson.  Unlike other scalar quartics, this scalar quartic is measurable at the LHC and for compressed spectra may even be the discovery channel for the stop or sbottom.  Observing this decay and measuring the value of the quartic interaction would give direct evidence that the underlying theory is supersymmetric as only a fermionic symmetry can relate a scalar quartic to a yukawa interaction.  This decay pathway is explored and situations where it can be the dominant decay mode for either the stop or the sbottom are highlighted.  Measurement of this decay mode at the LHC would unambiguously show that the underlying theory of nature was supersymmetric.  

\end{abstract}

\vspace{1cm}

\end{titlepage}

\section{Introduction}

Naturalness has been a guiding principle when looking for new physics at the LHC.  Finding experimental evidence for the cancelation of quadratic divergences is thus an important question.  Depending on how quadratic divergences are canceled, it can be very difficult to find any evidence of it at the LHC\cite{Hedri:2013ina}.  A slightly easier question is finding evidence for the symmetry that enforces naturalness\cite{Allanach:2010pp, Kane:2011tv, Perelstein:2003wd}.  Given the quantum numbers of new particles, there is a strong theory bias towards which of the many possible symmetries is the correct one; how the underlying structure of the symmetry is determined is an experimental question which should be addressed.

Supersymmetry is arguably the most elegant solution to the naturalness problem.  It is a fermionic symmetry which relates particles of different spin and their couplings.  It relates yukawa interactions, which dominate the phenomenology, to scalar quartic interactions, which are generally unobservable.  Proving that the underlying theory is supersymmetric is difficult.  Identifying the quantum numbers of the new particle is an important first step (see \cite{Wang:2008sw,Chang:2011jk} and references therein), but does not experimentally prove supersymmetry.  The gaugino sector can be shown to be supersymmetric by measuring the value of the gaugino's yukawa interactions\cite{Allanach:2010pp, Kane:2011tv} and showing that they are equal to gauge coupling constants.  The supersymmetric nature of the squark sector is much more difficult to determine (see e.g.~\cite{Blanke:2010cm}).

In this article, we focus on finding scalar quartic interactions. Scalar quartics are important for providing evidence that the squark sector is supersymmetric.  Only a fermionic symmetry can relate a yukawa interaction to a scalar quartic.  Finding this scalar quartic interaction and measuring its value would confirm the presence of a fermionic symmetry, which is by definition supersymmetry.

Most scalar quartics are unobservable at the LHC or cause three body decays that are strongly suppressed.  
The particular interaction of interest is the SUSY required quartic interaction\cite{Martin:1997ns}
\bea
\label{Eq: cross term}
\mathcal{L} &\supset& y_t y_b^* H_u \tilde t_R H_d^\dagger \tilde b_R^\dagger + h.c. = v y_t y_b^* H^-  \tilde t_R \tilde b_R^\dagger + \cdots
\eea
After electroweak symmetry breaking, the term shown in Eq.~\ref{Eq: cross term} gives a two body decay via the charged Higgs.  The decay sends a right handed stop (sbottom) to a right handed sbottom (stop) and a charged Higgs.  This decay is a LHC observable and related to the top and bottom yukawas by supersymmetry.  Observing this decay mode and measuring its value would unambiguously show that the squark sector is supersymmetric.

The sole decay resulting from Eq.~\ref{Eq: cross term} is a decay through a charged Higgs; there is no decay through a W boson because the couplings to the W boson are all a result of gauge invariance and the squarks involved are all right handed.  Alternatively one can also expand $H_u$ and $H_d$ in terms of both the physical and goldstone fields to see that the only three body interaction involves $H^+$.  Other terms such as the $\mu$ or A-terms induce decays through both the charged Higgs and the W boson.  Eq.~\ref{Eq: cross term} is unique in that it allows decays through the charged Higgs only.


Decays widths resulting from Eq.~\ref{Eq: cross term} are proportional to $y_t^2 y_b^2$ and are small compared to other decays.  Despite the small couplings, there are still  several natural scenarios where it dominates the decays.  When $\tan\beta$ is large, the coupling scales as $y_t^2 y_b^2 \sim \tan\beta^2$ and can dominate the decays.  Another natural scenario is if the right handed stop is the NLSP with a gravitino LSP and the right handed sbottom is the NNLSP, then the decay of the sbottom is mediated by Eq.~\ref{Eq: cross term}.  Finally, winos couple only to the left handed squarks.  If the wino is the LSP, the decay of the right handed sbottom and stop will again be dictated by the quartic.  Simplified models based on these cases will be considered in Sec.~\ref{Sec: collider}.

Light third generation squarks occur naturally in many scenarios such as gauge mediated SUSY breaking (see ~\cite{Giudice:1998bp} and references therein), gaugino mediation~\cite{Kaplan:1999ac,Chacko:1999mi} (due to suppressed scalar masses) and supersoft SUSY breaking~\cite{Fox:2002bu} (again due to suppressed scalar masses).  In fact, for models of the more minimal~\cite{Cohen:1996vb} or natural~\cite{Essig:2011qg,Brust:2011tb,Papucci:2011wy} supersymmetry, preventing the third generation squarks from obtaining negative mass squared due to two loop effects is a strong constraint~\cite{ArkaniHamed:1997ab}.  It is not difficult in these models to arrange for the lightest two MSSM fields to be the right handed sbottom and stop~\cite{Craig:2011yk,Auzzi:2011eu,Craig:2012di,Cohen:2012rm}.

The cascade through a charged Higgs is also interesting because a charged Higgs can be difficult to discover~\cite{Djouadi:2005gj,Basso:2012st}.  
For small masses, the most promising channel of discovery is via the decay of the top quark into a charged Higgs~\cite{Craig:2012pu,Craig:2013hca,Aoki:2011wd}.  Indirect bounds place a strong bound of $\sim 260$ GeV~\cite{Limosani:2009qg} on the mass of the charged Higgs.  At these masses, the charged Higgs decays into a top and bottom making it extremely difficult to discover due to large backgrounds.  
Even if the other scalars $A$ and $H$ are seen, observing the charged Higgs separately is needed to better understand the Higgs sector.
The decay channel of a stop(sbottom) to a charged Higgs and a sbottom (stop) is an opportunity to study the charged Higgs when it is heavy and/or difficult to discover.  

The rest of the article is arranged as follows.  Sec.~\ref{Sec: collider} sets up the simplified models that will be considered in this paper.  Sec.~\ref{Sec: competing decay modes} covers the competing channels and under what circumstances does the decay through the charged Higgs dominate.  Sec.~\ref{Sec: mixing} discusses the effect of left-right mixing on the results.  Sec.~\ref{Sec: measure} discusses under what conditions is measuring the value of the quartic interaction feasible.  Sec.~\ref{Sec: bounds} discusses the constraints from the LHC.  Finally, Sec.~\ref{Sec: conclusion} concludes.

\section{Simplified Models}
\label{Sec: collider}

Decays through charged Higgses is an important decay channel.  These decays are important to showing the existence of the operator
\bea
\label{Eq: search quartic}
\mathcal{L} &\supset& y_t y_b^* H_u \tilde t_R H_d^\dagger \tilde b_R^\dagger + h.c.
\eea
Discovery and subsequent measurement of this coupling would definitively prove that the squark sector was supersymmetric.

To explore the LHC's sensitivity to decays induced by the quartic Eq.~\ref{Eq: search quartic}, we motivate four simplified models of interest. The four models are differentiated by the identity of the LSP and which of the two squarks are heavier.  The first two are gauge mediation motivated scenarios while the second two are wino LSP motivated scenarios.

\subsection{Gravitino LSP Simplified Models}

The first two simplified models are gauge mediation motivated models.  In gauge mediation models, the LSP is always the gravitino while essentially any MSSM particle could be the NLSP.  Compared to the right handed sparticles, the left handed sparticles obtain an additional positive 1-loop RG contribution to their masses due to the wino.  Thus in the context of gauge mediation, it is natural to consider right handed stop or sbottom NLSPs~\cite{Kats:2011it}.  If the two lightest MSSM fields are the right handed stop and sbottom, we arrive at the simplified models of interest.

\begin{figure}
     \centering
     \includegraphics[width=0.45\linewidth]{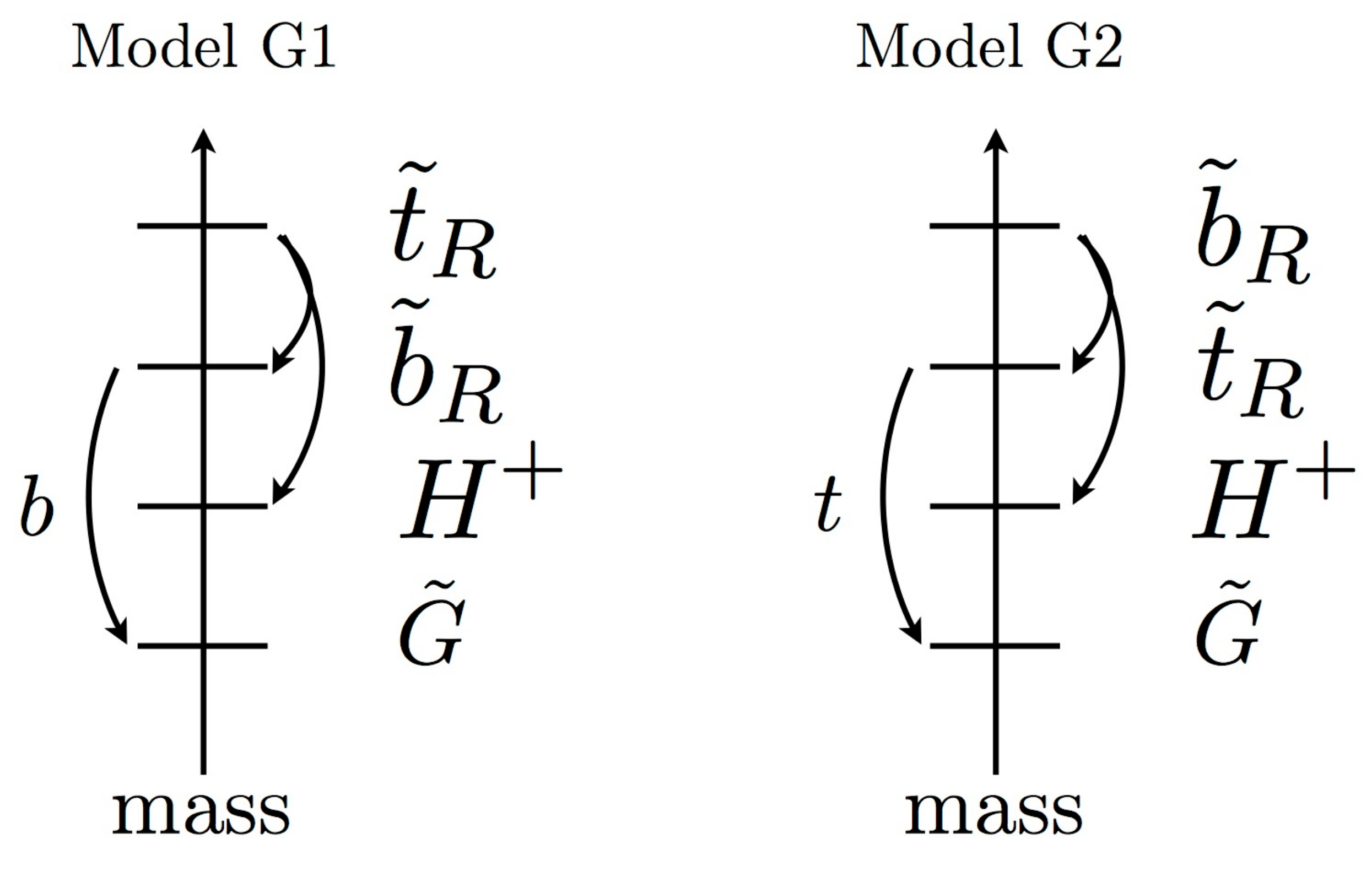}
     \caption{Mass spectrum and decay channels for the gauge mediation inspired simplified models G1 and G2.}  \label{Fig: mass spectrum}
\end{figure}

The models consist of a right handed stop, a right handed sbottom, a charged Higgs and a gravitino LSP.  The stop/sbottom decay to each other via a charged Higgs.  The simplified models will be referred to as models G1 and G2.  They are distinguished by which 3rd generation squark is heavier.  A visual representation of the simplified models and its decay chains are shown in Fig.~\ref{Fig: mass spectrum}.

The cascade of interest for model G1 is
\bea
\label{Eq: G2}
p p &\rightarrow& \tilde t_R + \bar \tilde t_R \rightarrow H^+ + H^- + \tilde b_R + \bar  \tilde b_R \\
\nonumber &\rightarrow& H^+ + H^- + b + \bar b + 2 \tilde G
\eea
while Model G2 has the cascade,
\bea
\label{Eq: G1}
p p &\rightarrow& \tilde b_R + \bar \tilde b_R \rightarrow H^+ + H^- + \tilde t_R + \bar  \tilde t_R \\
 \nonumber &\rightarrow& H^+ + H^- + t + \bar t + 2 \tilde G
\eea

To complete the description of the cascades, the decay of the charged Higgs must be specified.  
$H^+$ can decay to $W^+ + h_0$, $t + \bar b$, or $\tau^+ + \nu_\tau$.  The decay through a gauge boson vanishes in the decoupling limit.  LHC data indicates a Standard Model like Higgs boson~\cite{ATLAS:2013sla,CMS:yva} so that the decoupling limit is taken.  The remaining decays are into third generation fermions.  Direct production constraints from LEP require $m_{H^+} > 80$ GeV~\cite{Abbiendi:2013hk}.  For $H^+$ lighter than the top quark, there are direct detection bounds from the LHC due to top decays, see e.g. ~\cite{Aad:2012tj,CMS:mxa}.

   For small $\tan \beta$ and heavy $m_{H^+}$, $y_t$ is large so that the decay into tops and bottoms will dominate.
   If $\tan \beta$ is large, then the branching ratios into 3rd generation quarks versus leptons are determined by the bottom and tau masses so there is a 5.5:1 ratio of decay widths.  The simplest case is where $H^+$ is lighter than the top so that the only kinematically available channel is into the taus.

  Indirect detection bounds are more stringent but also more model dependent.  Constraints from $b \rightarrow s \gamma$ constrain $m_{H^+} > 260$ GeV~\cite{Limosani:2009qg}, but can be canceled by other light charged states~\cite{Garisto:1993jc}.  Models G1 and G2 have no other light charged states so we impose the bound $m_{H^+} > 260$ GeV.
  
  A heavy $H^+$ will decay predominantly into $t \bar b$ so in these simplified models we consider a 100\% branching ratio into $t \bar b$ for $H^+$.  Thus the final state of the cascade for Model G1 is 
\bea
2 t + 2 \bar t + b + \bar b + 2 \tilde G
\eea
while the final state for Model G2's cascade is
\bea
t + \bar t + 2 b + 2 \bar b + 2 \tilde G
\eea

\subsection{Wino LSP Simplified Models}

  If the LSP is a wino, then the right handed squarks only couple to it through mixing effects.  As the mixing is small, they will preferentially decay to each other before decaying into the wino as long.  The chargino component of the wino has two degrees of freedom as opposed to the single degree of freedom in the neutral component so that the branching ratio of the stop/sbottom to the chargino versus the neutralino is 2:1. 

\begin{figure}
     \centering
     \includegraphics[width=0.5\linewidth]{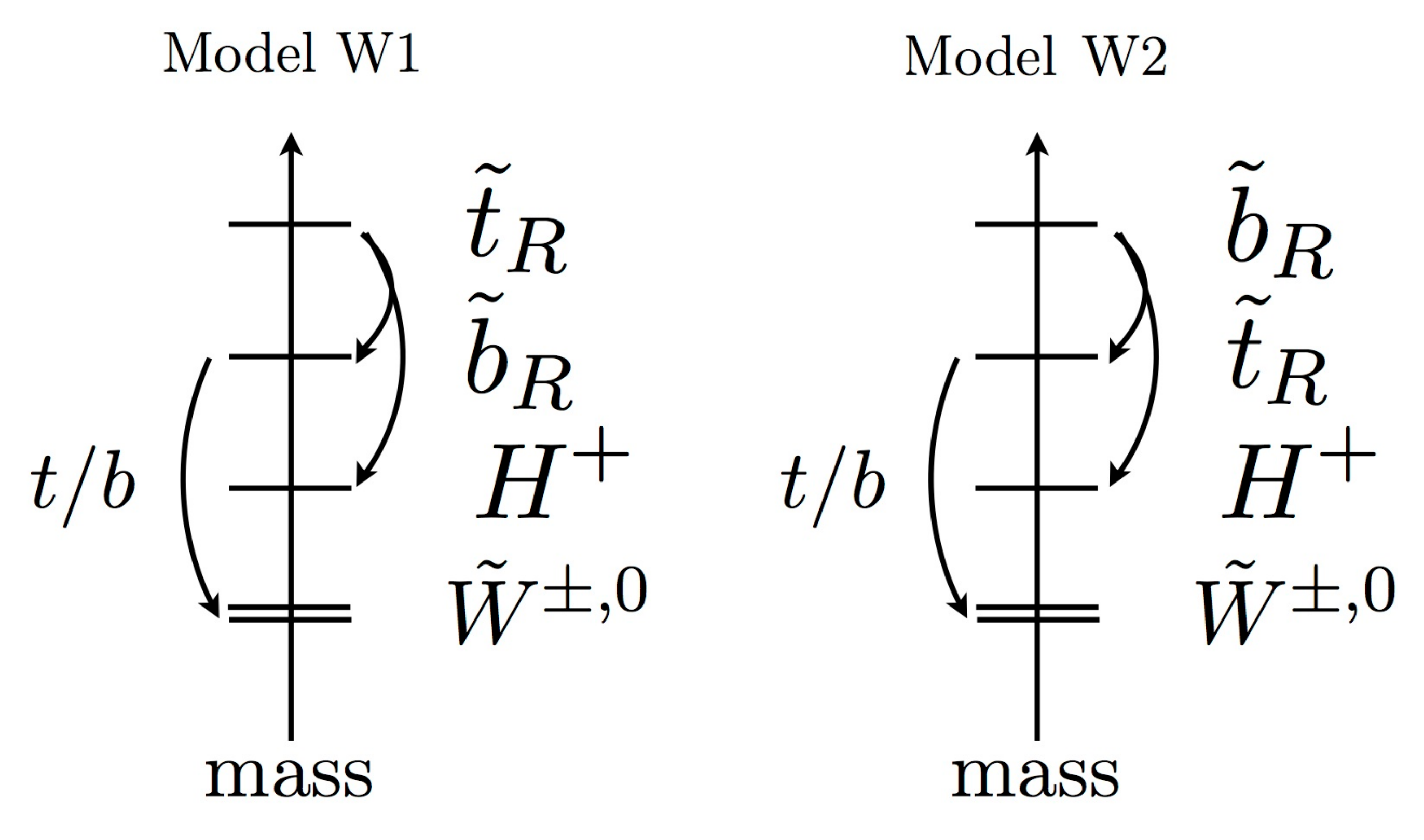}
     \caption{Mass spectrum and decay channels for the simplified models W1 and W2.}  \label{Fig: mass wino}
\end{figure}

The simplified models of interest in this section consist of a right handed stop, a right handed sbottom, a charged Higgs and a wino LSP with both chargino and neutralino components nearly degenerate.
  The cascades/models are shown visually in Fig.~\ref{Fig: mass wino}.  The first simplified model, Model W1, has a heavier stop so the cascades of interest are 
\bea
\label{Eq: W1}
p p &\rightarrow& \tilde t_R + \bar \tilde t_R \rightarrow H^+ + H^- + \tilde b_R + \bar  \tilde b_R \\ \nonumber
\tilde b_R &\rightarrow_\text{33\%}& b + \tilde W^0 \\ \nonumber
\tilde b_R &\rightarrow_\text{66\%}& t + \tilde W^- \rightarrow t + W^- + \tilde W^0 
\eea
The percentages listed are the branching ratios.  As the LSP is the wino with no mixing, $\tilde W^+$ and $\tilde W^0$ are roughly degenerate.  The chargino decays to the LSP through a highly off-shell W.

The second simplified model, model W2, has a heavier sbottom than stop so that the cascades of interest are 
\bea
\label{Eq: S2}
p p &\rightarrow& \tilde b_R + \bar \tilde b_R \rightarrow H^+ + H^- + \tilde t_R + \bar  \tilde t_R \\ \nonumber
\tilde t_R &\rightarrow_\text{33\%}& t + \tilde W^0 \\ \nonumber
\tilde t_R &\rightarrow_\text{66\%}& b + \tilde W^+ \rightarrow t + W^+ + \tilde W^0 
\eea
As before, the percentages listed are the branching ratios.

  As before, the charged Higgs decay modes are important.  For models W1 and W2, there are light charginos.  The wino contribution to $b \rightarrow s \gamma$ involves squarks as well as the chargino.  All of the left handed squarks have been decoupled from this simplified model, so the wino cannot be used to cancel the $H^+$ contribution.  Thus we place the same bound as before, $m_{H^+} > 260$ GeV.  The heavy $H^+$ decays predominantly into $t \bar b$ so in these simplified models, we consider a 100\% branching ratio into $t \bar b$ for $H^+$.    The final state of the cascade for models W1 and W2 are 
\bea
t + \bar t + b + \bar b + 2(t/b) + 2 \tilde W^0
\eea
with differing percentages of tops and bottoms.

\section{Competing decay modes}
\label{Sec: competing decay modes}

  This section considers various competing decay channels and explores parameter regions where the decay from the quartic 
\bea
\mathcal{L} &\supset& y_t y_b^* H_u \tilde t_R H_d^\dagger \tilde b_R^\dagger + h.c. \\
\nonumber &=& y_t y_b^* v H^- \tilde t_R \tilde b_R^\dagger + h.c. = \frac{m_t m_b}{v} (\tan\beta + \cot\beta) H^- \tilde t_R \tilde b_R^\dagger + h.c.
\eea  
  dominates.  It induces the decay $\tilde t_R (\tilde b_R) \rightarrow H^\pm + \tilde b_R (\tilde t_R)$ .   The results are briefly summarized in Tab.~\ref{Tab: channels}.  All branching ratio calculations in this section were done by taking tree level matrix elements and calculating decay widths analytically.   To simplify the number of parameters involved, we set $\tan\beta = 5$, the decaying squark mass is set to 1 TeV, and the decay products are taken to be massless.
  
\begin{table}
    \centering
    \begin{tabular}{cc}\toprule
        Competing channel (decay into) & Competing channel constitutes over 70\% of the branching ratio when \\
        \midrule
        bino & $\tan\beta \lesssim 75$ \\
        wino & mixing with left handed squarks $\gtrsim 0.05$ \\
        gluino & $\tan\beta \lesssim 200$ \\
        off-shell gluino & gluino mass less than 1.5 times the decaying squark's mass \\
        higgsinos & always for sbottoms. $\tan\beta \lesssim 160$ for stops \\
        gravitino & gravitino mass $\lesssim 4 \times 10^{-2}$ eV  \\
        $udd$ RPV & RPV coupling $\gtrsim 0.01$  \\
        \bottomrule
    \end{tabular}
    \caption{The status of various competing channels for the scalar quartic induced decay $\tilde t_R (\tilde b_R) \rightarrow H^\pm + \tilde b_R (\tilde t_R)$.  When not specified, $\tan\beta = 5$, the decaying squark mass was set to 1 TeV and the decay products were massless.} \label{Tab: channels}
\end{table}  
    
   \begin{figure}
     \centering
     \includegraphics[width=0.44\linewidth]{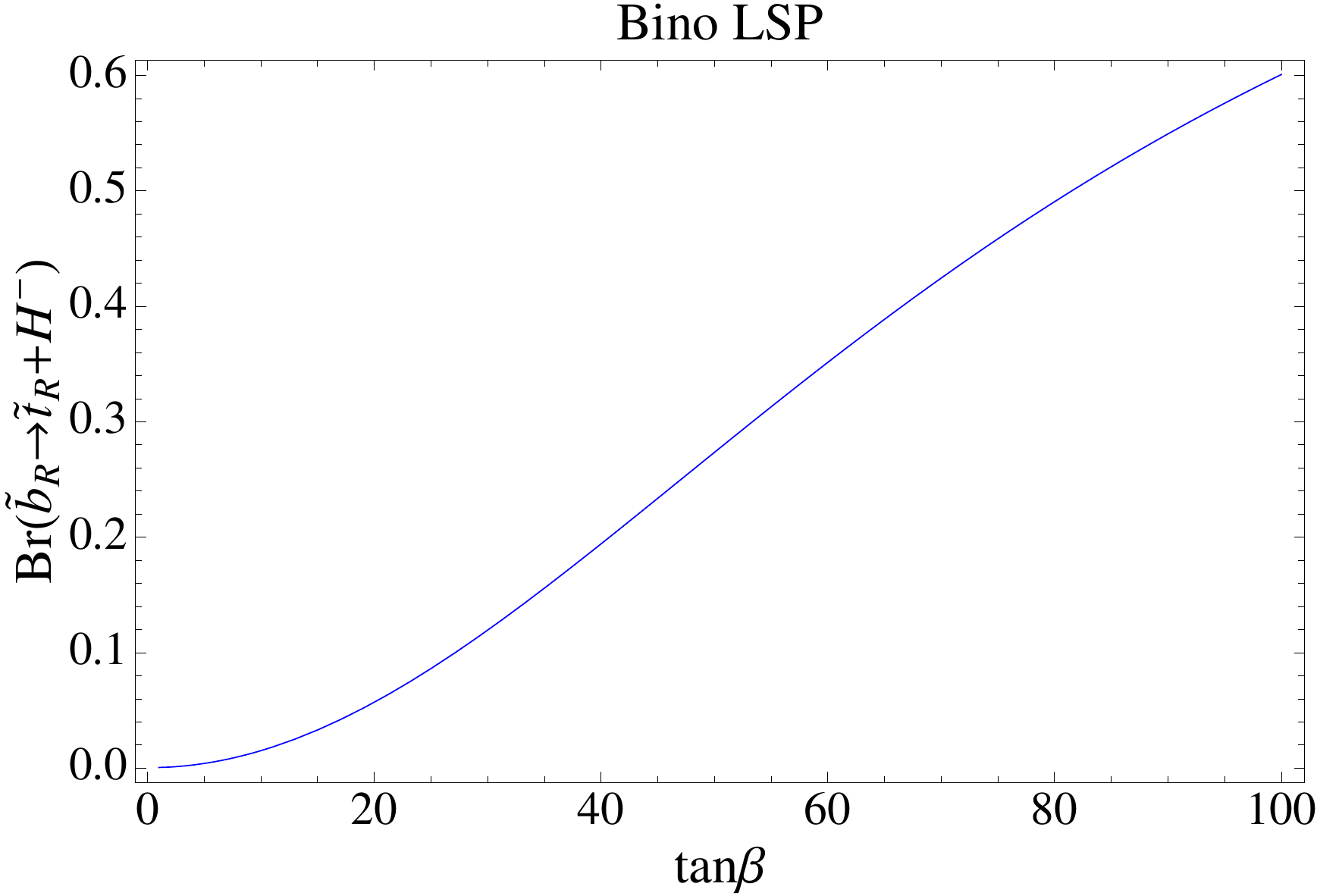}
     \includegraphics[width=0.45\linewidth]{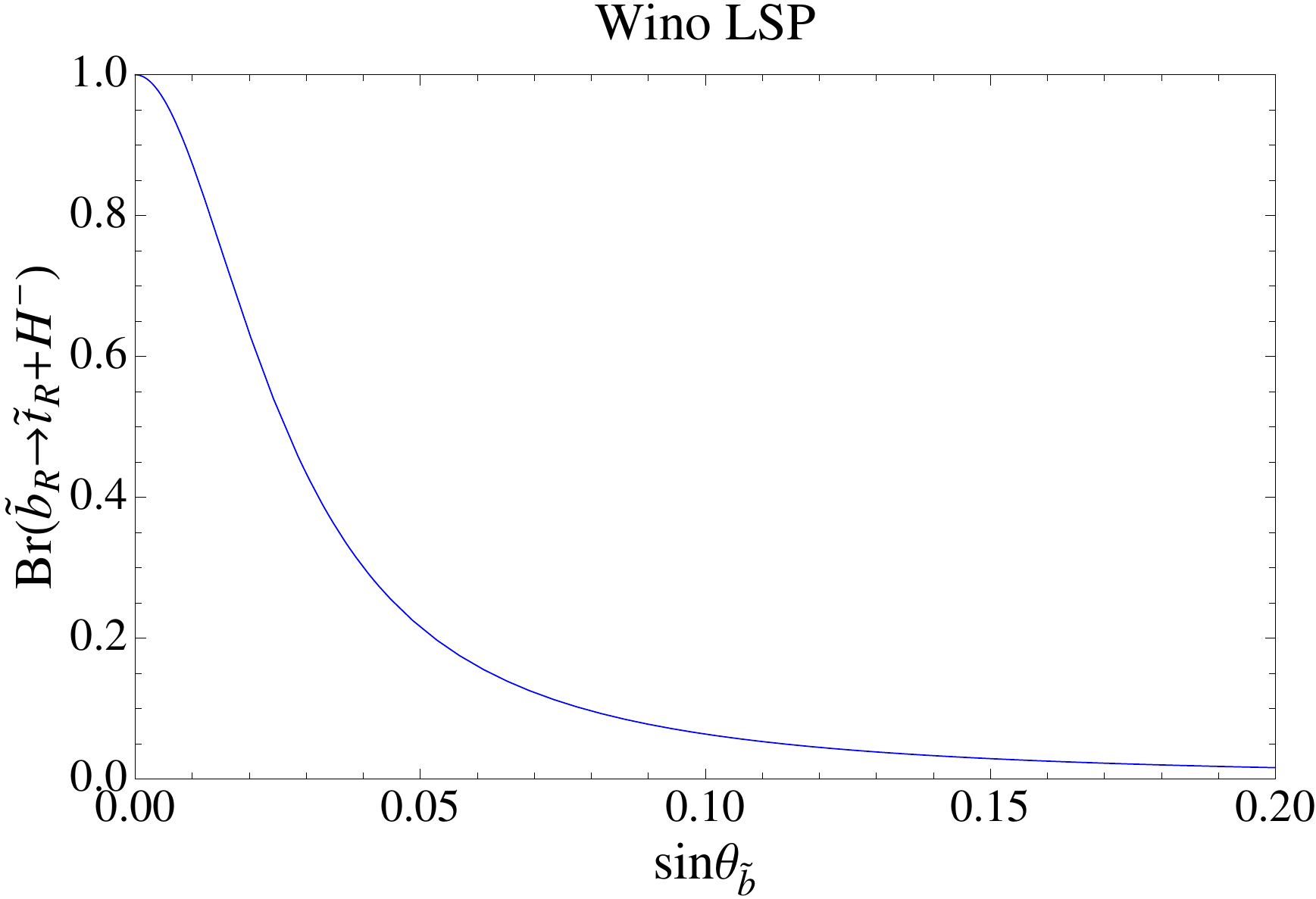}
     \caption{How the $\tilde b_R$ ($\tilde t_R$) branching ratios to $H^\mp + \tilde t_R (\tilde b_R)$ changes when there is a competing gaugino decay channel.  The decaying squark has its mass set to 1 TeV.  For the wino, $\tan\beta=5$ and the mixing angle between left and right handed sbottoms is varied.} \label{Fig: neutralinos}
 \end{figure}    
    
  The decay of a stop or sbottom to a gaugino and a quark occurs through a yukawa coupling of either gauge or yukawa strength.  These couplings give decay widths that scale as the mass of the stop (sbottom).  Thus at large squark masses, these decays will dominate.  The quartic interaction is linear in $\tan\beta$ so that large $\tan \beta$ will favor decays induced by the quartic.  The exception is the wino, which doesn't couple to the right handed squarks.
  
Fig.~\ref{Fig: neutralinos} shows how the branching ratio into the charged Higgses and the lighter stop/sbottom changes when a massless bino or wino is introduced.  Except for very large $\tan\beta$, the decay into the bino dominates.  As decays into winos proceeds only through mixing, $\tan\beta$ was set to five and the mixing angle was varied.  The value $\sin\theta_{\tilde b} = 0$ corresponds to a purely right handed sbottom.  The couplings of the squarks to the gluinos and higgsinos are larger than the coupling to the bino.  As a result, decays to the gluino and higgsinos are even more likely to dominate the decays when they are available.

\begin{figure}
     \centering
     \includegraphics[width=0.44\linewidth]{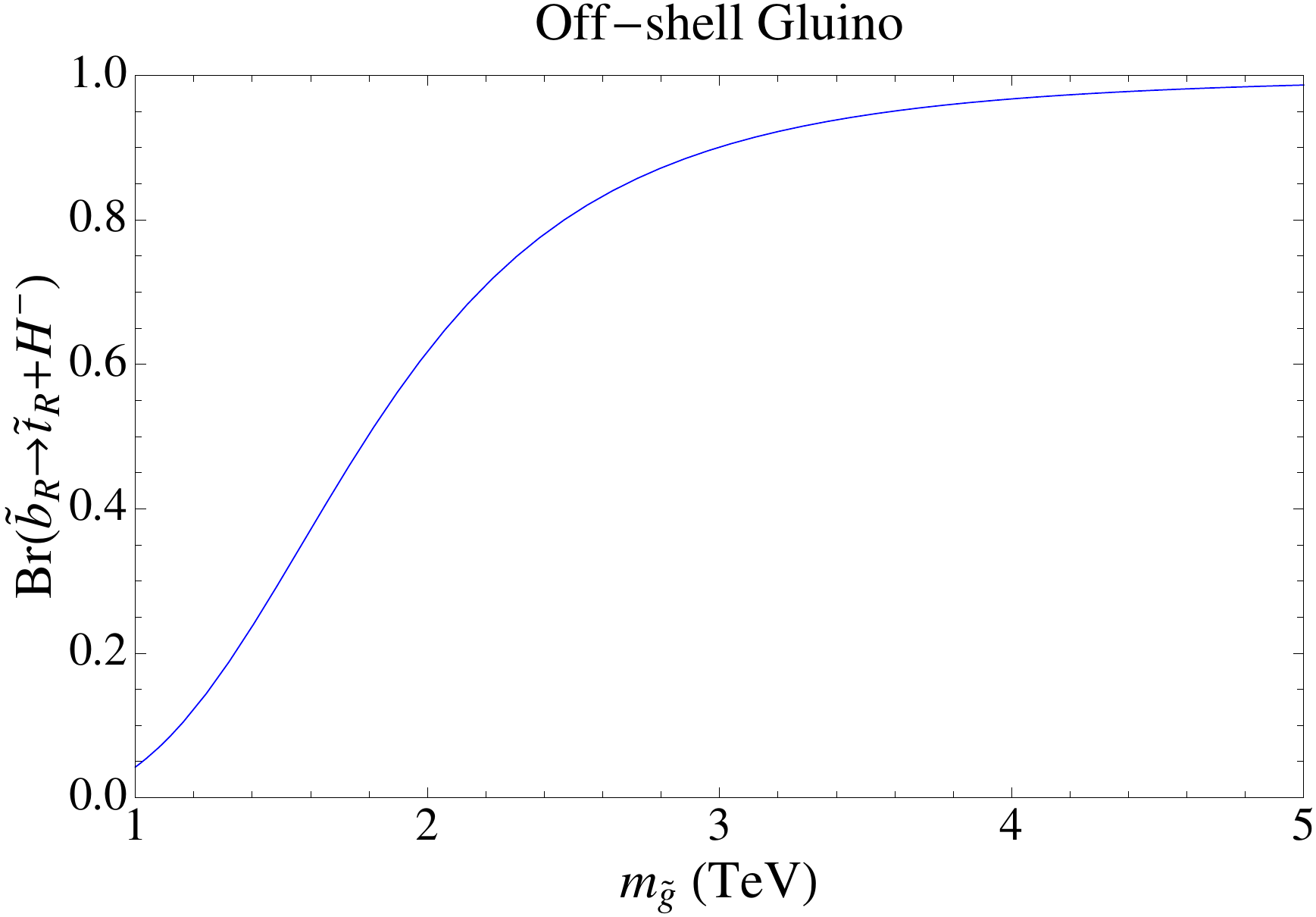}
     \includegraphics[width=0.45\linewidth]{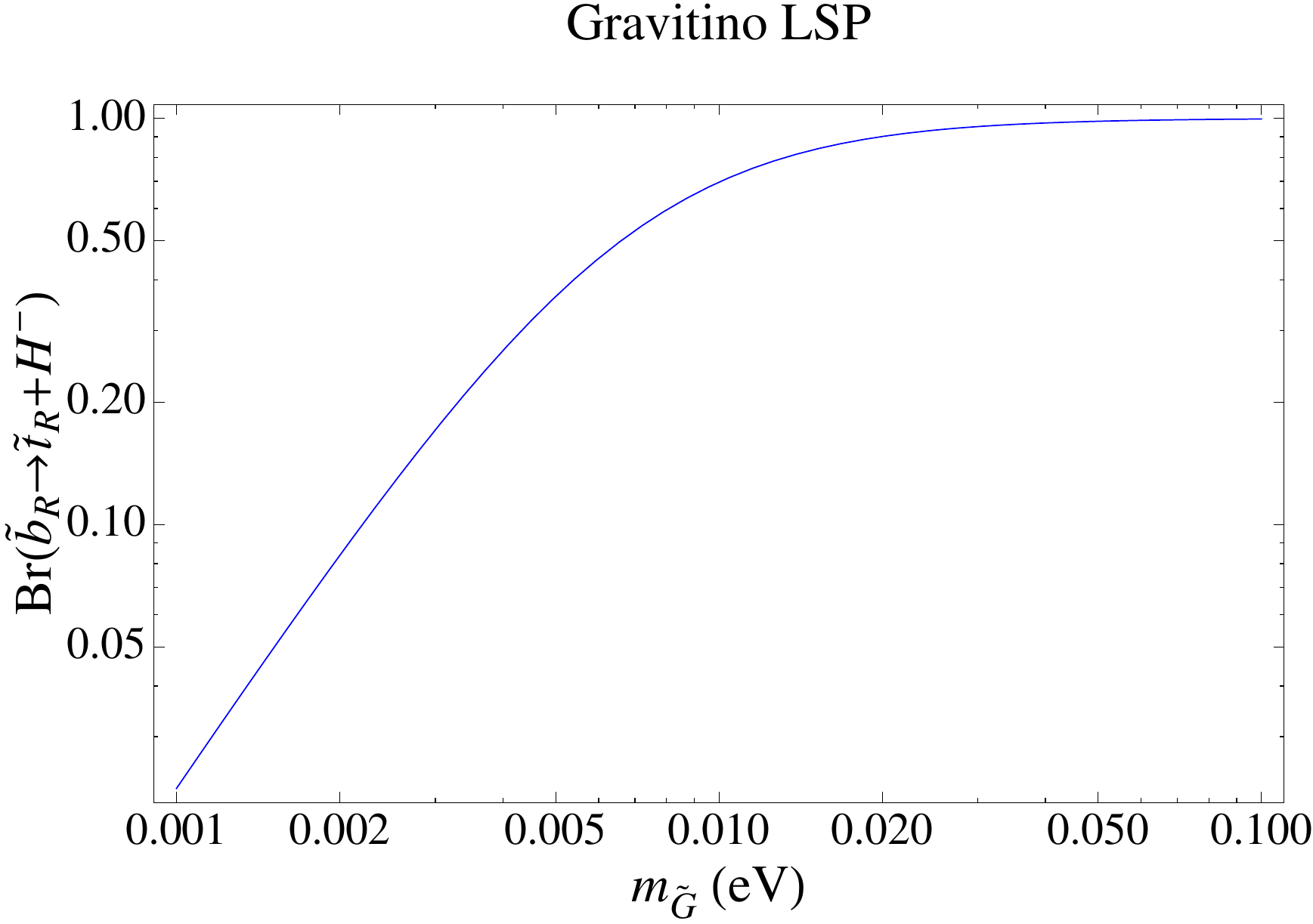}
     \caption{The dependence of the branching ratio of $\tilde b_R \rightarrow H^- + \tilde t_R$ on either the gravitino mass or the mass of an off-shell gluino.  We fix $\tan \beta=5$ and $m_{\tilde u_R}=1$ TeV.  }  \label{Fig: gluino}
 \end{figure}

Even when the stop or sbottom decays through an off-shell gaugino, the decay can still compete with the decay through a charged Higgs.  The most constrained gaugino is the gluino due to its large QCD sized couplings.  Depending on how heavy the gluino is, we can extrapolate between the two decay modes being dominant.  The three body decay has larger couplings but is suppressed by a phase space factor of about 1000 relative to the two body case.  The dependence of the branching ratio on the mass of the gluino is shown in Fig.~\ref{Fig: gluino}.
  
The direct decay to the gravitino could compete with the decay via $H^-$ depending on the mass of the gravitino.  The term in the Lagrangian coupling the goldstino\footnote{At energies much larger than its mass, the gravitino is well described by the goldstino.  Having spin 1/2 rather than 3/2, the goldstino is more familiar and easier to work with than the gravitino itself.} to $\tilde b_R$~\cite{Kats:2011it,Moroi:1995fs,Pradler:2007ne,Luo:2010he} is given by
\bea
\mathcal{L}_{\tilde b_R b \tilde G} &=& \frac{2 i}{F} \partial_\mu \tilde b_R \partial^\mu \bar \psi_{\tilde G} P_R b + h.c. \\
m_{\tilde G} &=& \frac{F}{\sqrt{3} M_{\text{pl}}}
\eea
The results are shown in Fig.~\ref{Fig: gluino}.

When the gravitino mass is roughly lighter than $10^{-2}$ eV, then its decay can compete with the scalar induced decay.  A $10^{-2}$ eV or lighter gravitino means a susy breaking scale of several TeV or less.  Most models have larger susy breaking scales, so there is a theory bias towards gravitinos heavier than $10^{-2}$ eV.

If there are RPV couplings, then squarks can decay directly into quarks.  We find that if the RPV coupling $\lambda$ is smaller than $10^{-3}$, then the decays through the charged Higgs will dominate over the RPV induced direct decays to quarks.

\section{Mixing effects}
\label{Sec: mixing}

Non-zero mixing is important for two reasons.  The first reason is that mixing with the left handed squarks opens up a new decay channel, the W boson.  We require that this new decay channel have a small branching ratio.  A second reason is that mixing will change the value of the $\tilde t_R \tilde b_R H^+$ coupling.  It will no longer receive all of its contributions from the scalar quartic interaction, but instead also receive contributions due to the $\mu$ and A-terms.  The focus of this paper is on the scalar quartic Eq.~\ref{Eq: cross term} not on the $\mu$ or A-terms.  Thus we study when the decay width comes from the scalar quartic rather than the $\mu$ or A-terms.

  Once electroweak symmetry is broken, there is mixing between the two squarks.  
For small mixing, the mixing angles scale schematically as
\bea
\sin \theta_{\tilde t} &\sim& \frac{(\mu \cot\beta+ A_t) m_t}{m_{\tilde Q}^2-m_{\tilde t_R}^2} \\
\sin \theta_{\tilde b} &\sim& \frac{(\mu \tan\beta + A_b) m_b}{m_{\tilde Q}^2-m_{\tilde b_R}^2}
\eea
Decays through W bosons occur through purely left handed particles and so only occur after both the sbottom and stop have mixed.  The effect of mixing on the decay width involves only a single mixing and so the dominant effect of mixing is to change the value of the decay width.

The effect of mixing on the decay width of $\tilde b_R$ is suppressed by $\mu^2/(m_Q^2-m_{\tilde q}^2)$ while the suppression of the decay via a $W^+$ is at least that much and for large $\tan\beta$ is $m_b^2 \mu^2 / m_Q^4$.  Requiring that mixing effects do not change the total width by more than 10\% gives the weak constraint that $m_Q$ is only about a factor of a few heavier than $\mu$ and $m_{\tilde q}$.

\begin{figure}
     \centering
     \includegraphics[width=0.45\linewidth]{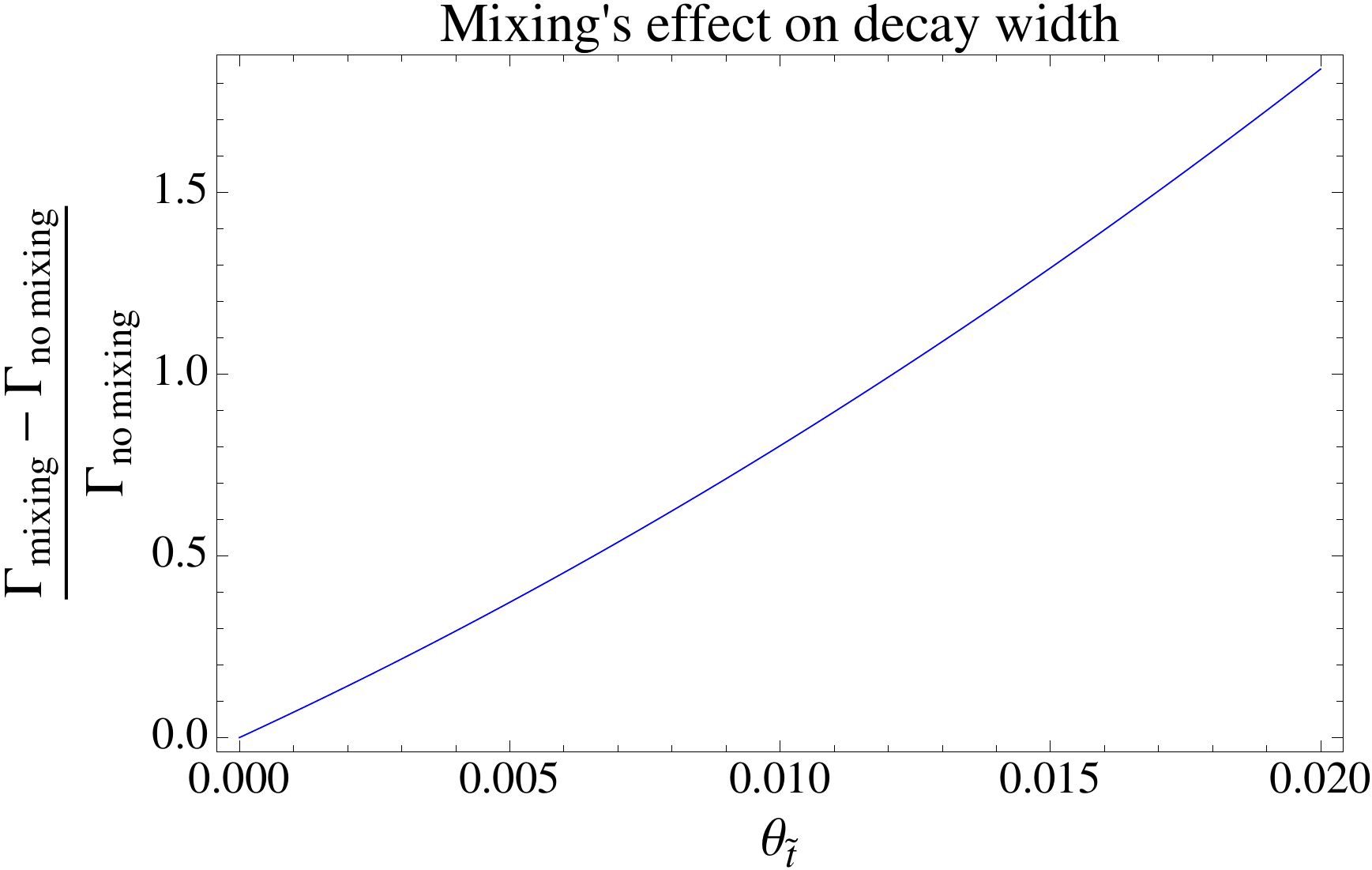}
     \includegraphics[width=0.425\linewidth]{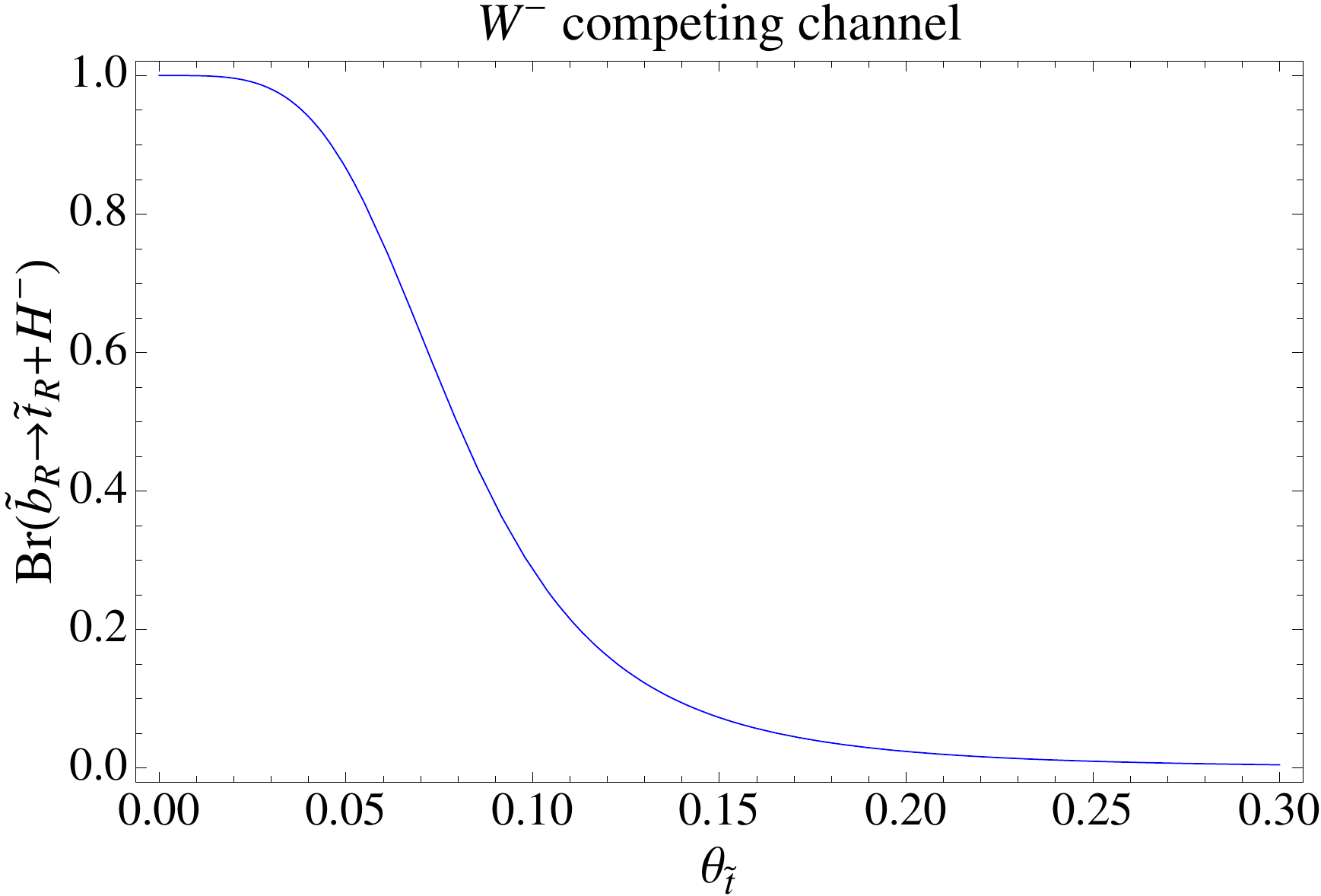}
     \caption{The effect of mixing on the additional decay channel through a W boson and the dependence of the decay width on mixing angles.  Large fractional changes in the decay width indicate that the decay width is no longer coming the scalar quartic.  We fix $A_t = A_b = 0$, $\mu = 1200$ GeV, $\tan\beta = 5$.  The mixing angles are not completely independent.  When varying $\theta_{\tilde t}$, we enforce the relationship $\theta_{\tilde t} \sim \frac{m_t}{\tan^2\beta m_b}\theta_{\tilde b}$.}  \label{Fig: mixings}
\end{figure}

To quantify the effects of mixing, we look at the fractional change in the decay width ($\frac{\Gamma_{\text{mixing}}-\Gamma_{\text{no mixing}}}{\Gamma_{\text{no mixing}}}$) while holding $A_{t,b}=0$, $\mu = 1200$ GeV, and $\tan\beta = 5$.  The stop and sbottom mixing angles are related to each other roughly by $\theta_{\tilde t} \sim \frac{m_t}{\tan^2\beta m_b}\theta_{\tilde b}$ so we keep this relationship and vary $\theta_{\tilde t}$.  The results are shown in Fig.~\ref{Fig: mixings}.

After $\theta_{\tilde t} \gtrsim 10^{-3}$, the decay width becomes significantly changed.  Note that small mixing angles occur naturally due to the small quark masses and $\tan\beta$ effects.  $m_Q$ in the few TeV range is enough to achieve these small mixing angles.

If the LHC were to measure the coupling and see that it agreed with the predictions from the scalar quartic, it would be direct evidence of supersymmetry.  However, if it measured the coupling and saw it to be too large, it is not evidence against supersymmetry as there may simply be nontrivial mixing effects.

%

\section{Measuring the quartic}
\label{Sec: measure}

To demonstrate that the squark sector is supersymmetric, one must measure the value of the quartic coupling inducing the decay.  There are two avenues for measuring the quartic coupling.  The first is to directly measure the decay width.  This measurement is very difficult and not feasible at the LHC.  The second is if there are competing channels.  If the coupling of the competing channel to the squarks is known by some other means, then the relative branching ratios can be used to determine the value of the quartic coupling.

  There are two promising situations.  The first is a bino LSP at large $\tan\beta$.  To have a least a 20\% branching ratio through a charged Higgs, we need $\tan\beta > 60$.  $\tan\beta$ can be measured from the Higgs sector.  Associated production of a bino and a squark can be measured to establish the couplings of the bino to squarks.  Finally a measurement of the quartic coupling could be made from the relative branching ratio of decays through $H^+$ versus direct decays to the bino.

Large $\tan\beta$ implies large couplings between the Higgs sector and the bottom quarks.  At large $\tan\beta$, the production of the neutral Higgses occurs through gluon fusion and associated b production.  Increasing $\tan\beta$ to large values increases the coupling to the bottom quarks and hence the production cross section.  The decay of $H$ and $A$ to $\mu^+ \mu^-$ and $\tau^+ \tau^-$ can be probed.  For a $\tan\beta$ of 60, $m_A > 500$ GeV~\cite{Aad:2012yfa}.  At tree level, the charged Higgs mass is $m_{H^+}^2 = m_A^2 + m_W^2$ leading to a bound of also about 500 GeV for $m_{H^+}$.  Allowing the squarks to decay through an on-shell charged Higgs puts them at masses around 1200 GeV or larger.  

 The second possibility is a decay via an off-shell gluino.  As before, $\tan\beta$ can be measured from the Higgs sector and associated production can be used to measure the coupling to the stop/sbottom.  The relative branching ratios then provide a check of susy.  
 
 A difficulty arrises when the squark decays via an off-shell gluino; it could potentially give the same final states as a decay via a charged Higgs.  If the charged Higgs is heavier than the top mass, its dominant decay mode is to a top and bottom resulting in the two cascades having the same final states.  
   The simpler alternative is that the charged Higgs has a mass lighter than the top mass so that it decays primarily into $\tau + \nu_\tau$.  The final scenario for competition from an off-shell gluino are right handed stop/sbottom beyond the current bounds, a gluino a little heavier than the stop/sbottom, and a gravitino LSP.  The two competing decay channels are
\bea
p p &\rightarrow& \tilde b_R + \bar \tilde b_R \rightarrow t + \bar t + b + \bar b + \tilde t_R + \bar  \tilde t_R \rightarrow 2 t + 2 \bar t + b + \bar b + \slashed{E}_T\\ 
p p &\rightarrow& \tilde b_R + \bar \tilde b_R \rightarrow H^+ + H^- + \tilde t_R + \bar  \tilde t_R \\
\nonumber &\rightarrow& \tau^+ + \tau^- + \nu_\tau + \bar \nu_\tau + t + \bar t + \slashed{E}_T
\eea
A similar cascade would occur if the sbottom were lighter than the stop.  A comprehensive study of these scenarios will be left for future work.

Finally competition from RPV or gravitinos is untenable because establishing the mass of the gravitino or the value of the RPV coupling would be needed.  Measuring these values usually require displaced vertices.  In these situations, decays through $H^+$ would dominate so there is no competing channel.  A significant branching ratio into winos involves large mixing.  As shown in Fig.~\ref{Fig: mixings}, in these cases, the decay width comes more from the mixing effects than the scalar quartic.

\section{LHC bounds on the Simplified Models}
\label{Sec: bounds}

This section applies ATLAS searches to the simplified models specified in Sec.~\ref{Sec: collider}.  The models G1,G2, W1 and W2 are all characterized by $m_{\tilde b_R}$, $m_{\tilde t_R}$, $m_{H^+}$ and $m_{LSP}$.  A four dimensional parameter space is too large to scan effectively so we set
\bea
m_{H^+} = 300 GeV \quad m_H = m_L + 350 \, \text{GeV}
\eea
where $m_H$ ($m_L$) is the mass of the heavier (lighter) squark.  The LSP and NLSP mass are then scanned over in increments of 25 GeV.

These cascade decay have many tops and bottom quarks leading to a large number of b jets and leptons.  These decays are very similar to the decay topology $\tilde g \rightarrow 2 (t/b) + \chi_0$ except with 2 additional 2-bjets.  To demonstrate the LHC's sensitivity, we consider the ATLAS search for  $\tilde g \rightarrow 2 (t/b) + \chi_0$~\cite{atlas2013061}.  The bounds from this search on the cascade are compared to the ATLAS bounds on the direct search of the NLSP.  Details of event generation and validation are given in App.~\ref{App: details}.

The ATLAS gluino search has 9 different search regions with both 0 and 1+ lepton search regions, all of which require at least 3 b-jets.  Bounds are placed by calculating the signal contribution to each of the 9 signal regions and simply observing when it crosses the 95\% confidence level limits.  No combination of the different signal regions is used.

  Most of the search regions are similar in sensitivity.  Given the presence of leptons, it is not surprising that the most constraining search regions involve leptons.  The most constraining search region was SR-1l-6j-B.  The important cuts are 6 or more jets, three or more b jets, $\slashed{E}_T > 225$ GeV, $m_{eff} > 800$ GeV and $\slashed{E}_T / \sqrt{H_T} > 5 \sqrt{\text{GeV}}$.

 \begin{figure}
     \centering
     \includegraphics[width=0.43\linewidth]{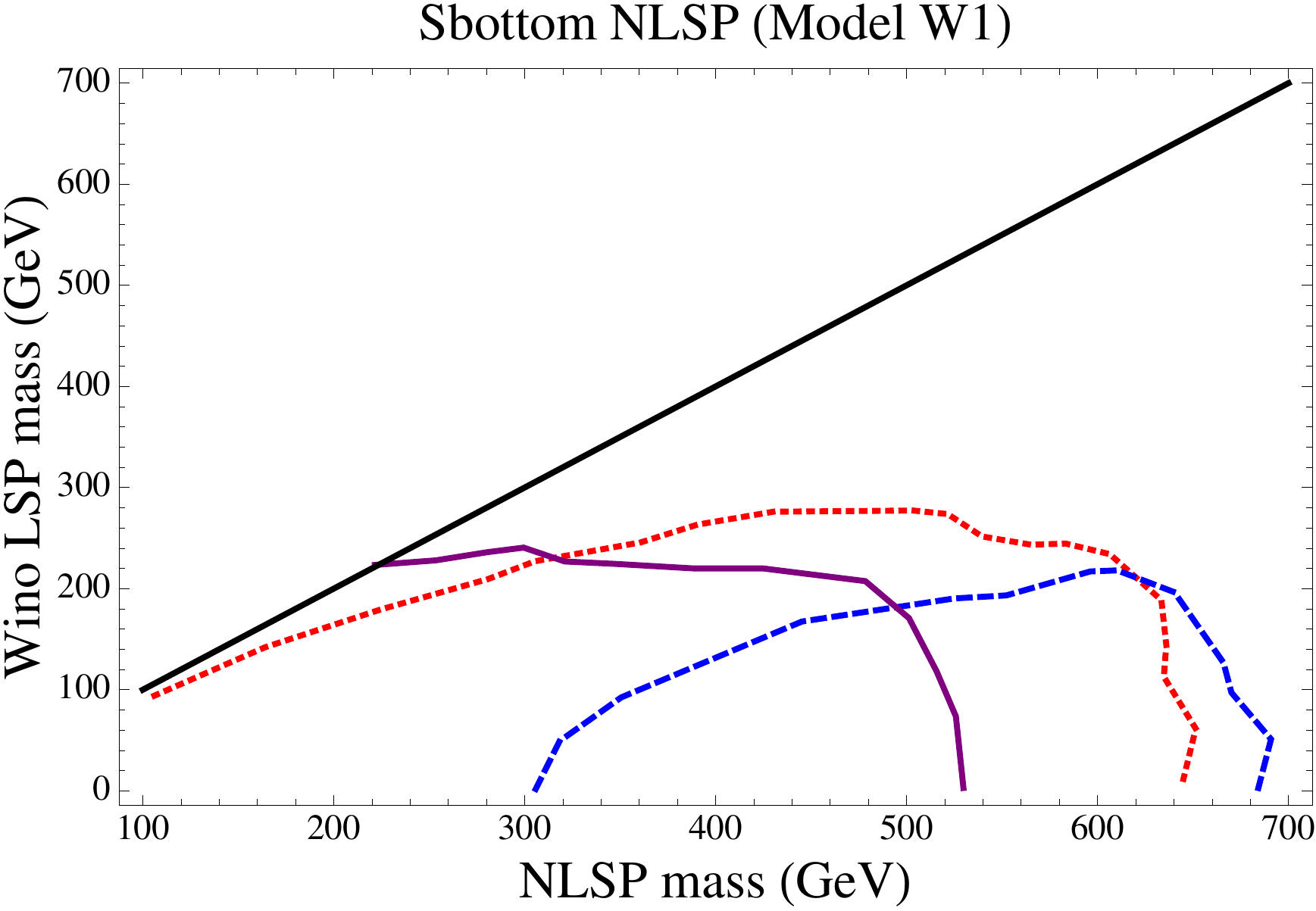}
     \includegraphics[width=0.43\linewidth]{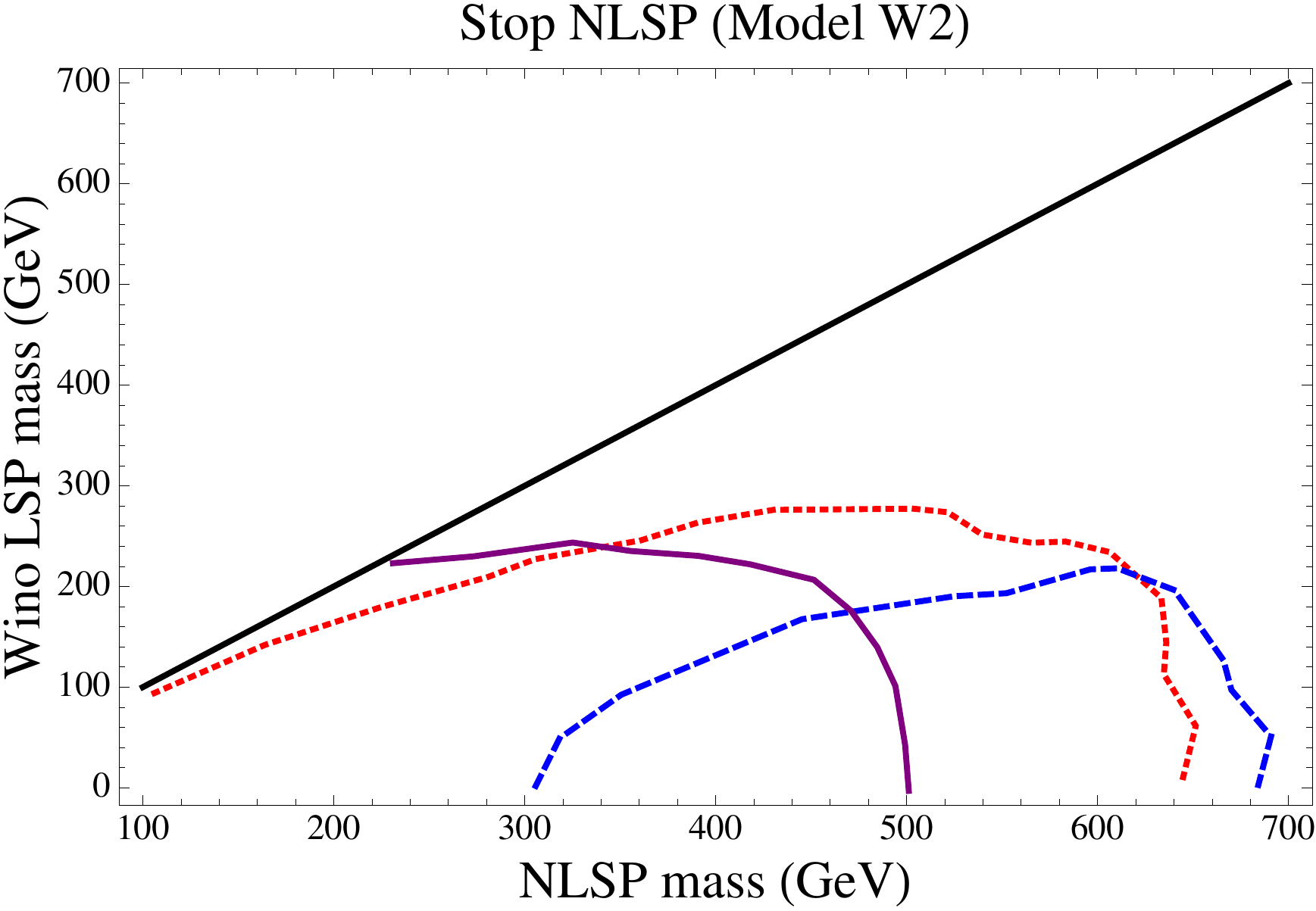}
     \includegraphics[width=0.43\linewidth]{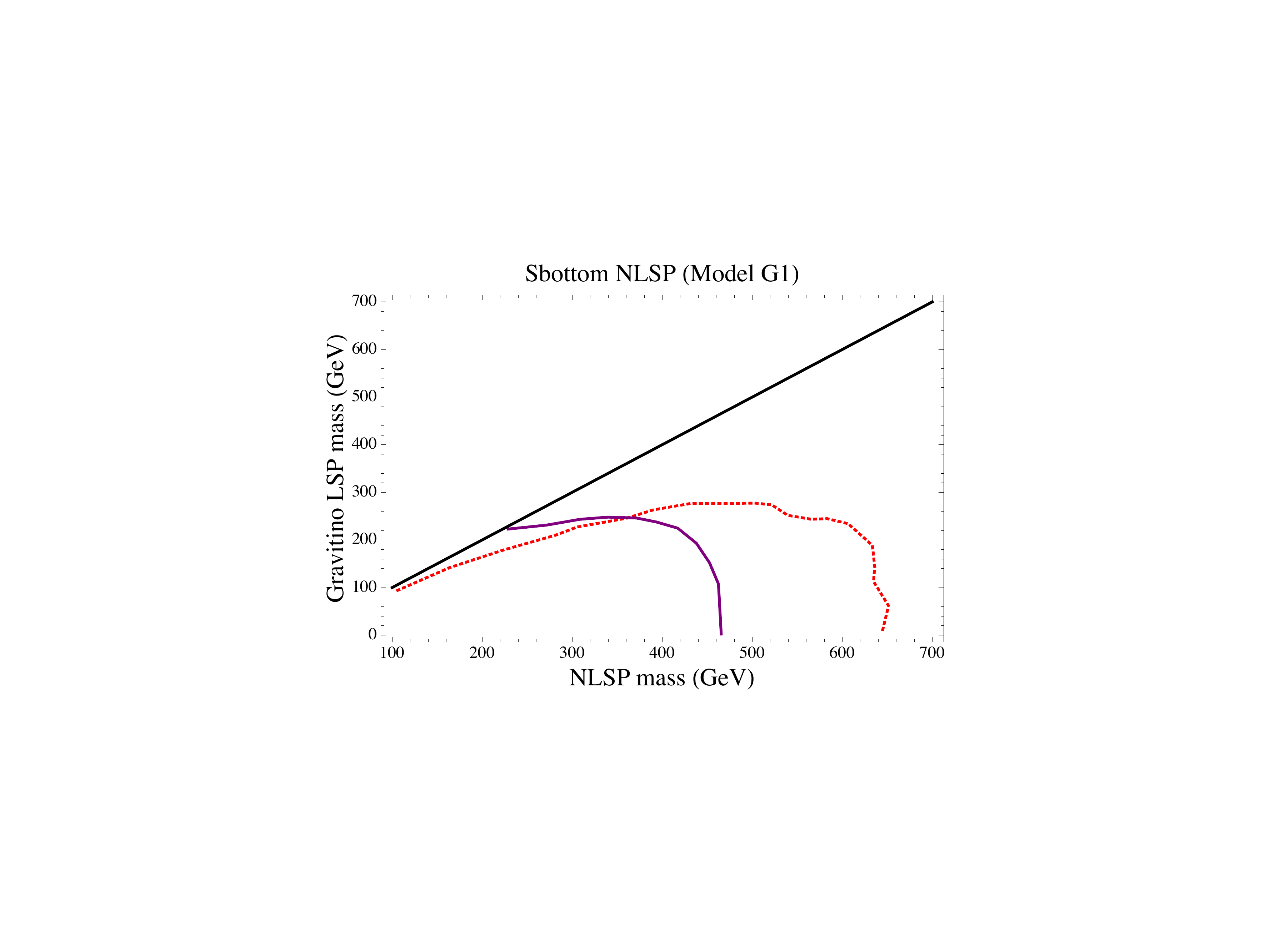}
     \includegraphics[width=0.44\linewidth]{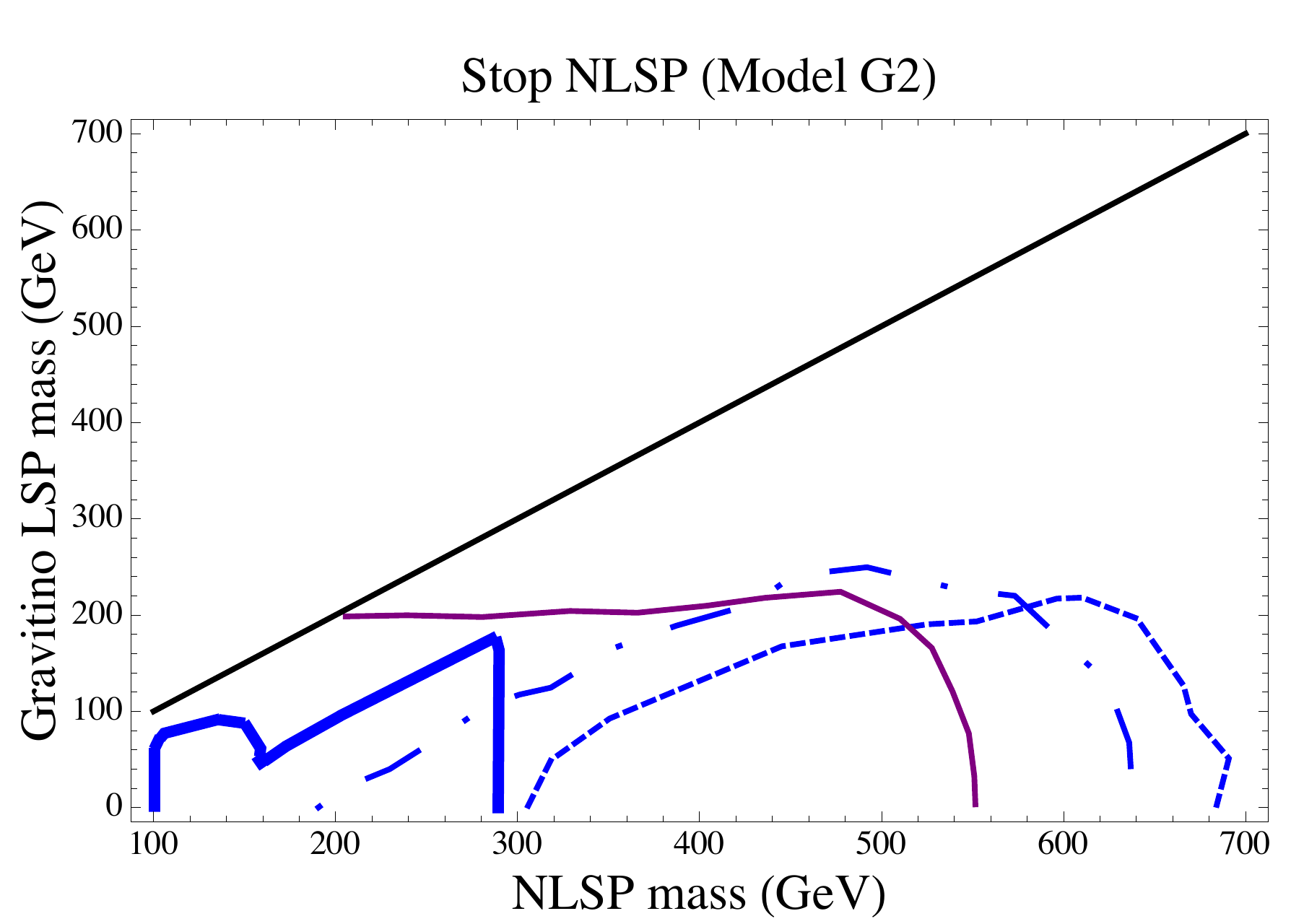}
     \caption{95\% confidence level bounds on the NLSP/LSP parameter space for the simplified models discussed in Sec.~\ref{Sec: collider}.  The heavier squark is 350 GeV more massive than the lighter squark and the charged Higgs has a mass of 300 GeV.  Bounds on the cascade are shown in solid purple while the bounds on the sbottom/stop NLSPs are shown in dotted red/blue (dashed, line-dot-line, and thick solid depending on the search).  We see that current searches already place bounds on the cascade that are comparable with searches for the direct production of the lighter squark.  There are regions of parameter space where the cascade would be seen before the direct production of the NLSP.}  \label{Fig: Bounds}
\end{figure}

The exclusions are shown in Fig.~\ref{Fig: Bounds}.  The solid black line shows when the LSP and the NLSP squark have the same mass.  The dotted red (blue) line shows the 95\% confidence level bounds on a sbottom (stop) decaying 100\% to an LSP and a bottom (top) coming from Ref.~\cite{Aad:2013ija}  (blue dashed line~\cite{ATLAS:2013cma}, blue line-dot-line~\cite{ATLAS:2013pla}, thick blue solid line~\cite{atlas2013065}).  Finally, the constraints on the cascade coming from \cite{atlas2013061} are shown with a solid purple line.  For the wino LSP, the decay channels of the NLSP are in a 2:1 ratio to the chargino:neutralino.  As neither bound strictly applies, both bounds are shown.  

  We see for both simplified models that the bounds on the cascades are non-negligible.  In most of parameter space, discovery of the cascade would happen shortly after the discovery of the direct searches
for the NLSP.  In squeezed regions, the cascade is seen first.  This effect is enhanced when the NLSP decays primarily through a top. 

\section{Conclusion} \label{Sec: conclusion}

In this article, we discussed a decay mode that results from a supersymmetric scalar quartic.  Discovery of this decay mode and its possible measurement would show the existence of a scalar quartic interaction which is related to a yukawa interaction by a symmetry.  Finding this relationship would prove that our underlying theory is supersymmetric.  This quartic interaction is $H_u H_d^\dagger \tilde t_R \tilde b_R^\dagger$ and gives decays via a charged Higgs.  

As a scalar interaction suppressed by yukawa interactions, this decay mode will not dominate except for very specific circumstances.  The gauginos have to be sufficiently heavier than the stops and sbottoms and the gravitino cannot be too light.  For R-parity conserving cases, the LSP can be the gravitino, the wino (for small mixing) or the bino (for large $\tan\beta$).  For RPV cases the magnitude of the RPV couplings must be sufficiently small.

We introduced four simplified models designed to capture the simplest cases where the decays are set by the scalar quartic.  These four models have wino or gravitino LSPs and have cascades involving charged Higgses.  Collider searches place non-trivial bounds on cascades and in some instances, the cascades may even be seen before the direct production of the NLSP.

\section*{Acknowledgements}
AH would like to thank Nima Arkani-Hamed and Nathaniel Craig for helpful discussion.  AH would like to thank Nathaniel Craig, Eder Izaguirre, Mariangela Lisanti and a JHEP referee for helpful comments on the draft.  AH is supported by the US DOE under contract number DE-SC0009988.

\appendix
\section{Details of Event Generation and Validation}
\label{App: details}

The events used were generated in  {\tt MadGraph 1.5.6} ~\cite{Alwall:2011uj} and showered in {\tt Pythia 6.4.14} ~\cite{Sjostrand:2006za}.  Events were matched with up to 2 additional jets using the MLM-matching scheme implemented in {\tt MadGraph}.  After showering, all hadron-level events are passed through a ``theorist's'' detector.  The search that this paper considers is ATLAS's $\tilde g \rightarrow 2 t + \chi_0$ search~\cite{atlas2013061} so this ``theorist's'' detector was designed using the same object reconstruction as described in the paper.

In more detail, the electrons (muons) were required to satisfy isolation requirements.  The total $p_T$ of all charged objects within a small cone $\Delta R = 0.2$ around the lepton were required to be smaller than 16\%(12\%) of the electron(muon) $p_T$.
  The {\tt Pythia} output was clustered into $0.1 \times 0.1$ cells in $\eta\!-\!\phi$ space and each cell was normalized to a massless four-vector pseudoparticle.  The events were then clustered using {\tt FastJet 3.0.3} ~\cite{Cacciari:2011ma} into R=0.4 anti-kt jets.  Overlap between jets and leptons were resolved by throwing away jets within $\Delta R = 0.2$ of an electron and then removing leptons within $\Delta R < 0.4$ of remaining jets.

  In the ATLAS search, b-tagging was done by a neural-network.  In this work, to simulate the neural-network, a jet was tagged as a parton level b-jet if there was a parton level b quark within $\Delta R < 0.3$ of the jet.  The efficiency versus jet $p_T$ curves for the 70\% efficiency working point of Ref.~\cite{ATLAS:2012ima} were modeled and used for tagging b-jets.  Uniform rejection factors of 137, 5 and 13 were used for light quarks, c-quarks and $\tau$ leptons.  Even after the efficiencies, the tagging of b-jets was still too efficient compared to the validation sample so a uniform additional efficiency of 90\% was applied to match the ATLAS data.

Event generation and search code was validated by generating gluino pair production with $m_{\tilde g} = 1300$ GeV decaying either through a pair of tops or bottoms to a 100 GeV LSP .  Histograms were compared and the results were found to match to within 20\%.  The cut flow tables were also reproduced and are shown in Tab.~\ref{Tab: tt1} and \ref{Tab: tt2}.  We notice that the ATLAS events have less jets than our Madgraph events.  The lepton identification is also better for ATLAS than it is for our toy detector.  In the validation sample, these point in the opposite directions and approximately cancel each other out.

\begin{table}
    \centering
    \begin{tabular}{ccccc}\toprule
        \multicolumn{5}{c}{$\tilde g \rightarrow t \bar t \chi_0$}\\
        \midrule
        \multicolumn{2}{c}{0-lepton Selection Criteria} & \multicolumn{3}{c}{Absolute Efficiency (ATLAS/Ours)}\\
        \midrule
        \multicolumn{2}{c}{$\ge$ 4 jets with $p_T > 30$ GeV} & \multicolumn{3}{c}{96.9/99.0\%}\\
        \multicolumn{2}{c}{First jet $p_T > 90$ GeV} & \multicolumn{3}{c}{96.9/98.4\%}\\
        \multicolumn{2}{c}{$\slashed{E}_T > 150$ GeV} & \multicolumn{3}{c}{88.3/89.1\%}\\
        \multicolumn{2}{c}{Lepton veto} & \multicolumn{3}{c}{45.9/49.5\%}\\
        \multicolumn{2}{c}{$\delta \phi_{min}^{4j} > 0.5$} & \multicolumn{3}{c}{30.0/34.7\%}\\
        \multicolumn{2}{c}{$\slashed{E}_T/m_{eff}^{4j} > 0.2$} & \multicolumn{3}{c}{25.9/30.3\%}\\
        \multicolumn{2}{c}{$\ge 7$ jets with $p_T > 30$ GeV} & \multicolumn{3}{c}{24.6/25.3\%}\\
        \multicolumn{2}{c}{$\ge 3$ b-jets with $p_T > 30$ GeV} & \multicolumn{3}{c}{11.5/12.2\%}\\
        \midrule
        && SR-0l-7j-A & SR-0l-7j-B &  SR-0l-7j-C\\
        \multicolumn{2}{c}{$\slashed{E}_T > 200-350-250$ GeV} & 11.3/12.0\% & 9.2/9.6\% & 10.8/11.5\% \\
        \multicolumn{2}{c}{$m_{eff}^{incl} > 1000-1000-1500$ GeV} & 11.3/11.9\% & 9.2/9.6\% & 9.5/8.9\% \\
        \bottomrule
    \end{tabular}
    \caption{A comparison of 0 lepton cut-flow tables between ATLAS and our ``theorist's'' detector for gluinos decaying through a pair of top quarks.} \label{Tab: tt1}
\end{table}

\begin{table}
    \centering
    \begin{tabular}{ccccc}\toprule
        \multicolumn{5}{c}{$\tilde g \rightarrow t \bar t \chi_0$}\\
        \midrule
        \multicolumn{2}{c}{1-lepton Selection Criteria} & \multicolumn{3}{c}{Absolute Efficiency (ATLAS/Ours)}\\
        \midrule
        \multicolumn{2}{c}{$\ge$ 4 jets with $p_T > 30$ GeV} & \multicolumn{3}{c}{96.9/99.0\%}\\
        \multicolumn{2}{c}{First jet $p_T > 90$ GeV} & \multicolumn{3}{c}{96.8/98.4\%}\\
        \multicolumn{2}{c}{$\slashed{E}_T > 150$ GeV} & \multicolumn{3}{c}{88.3/89.1\%}\\
        \multicolumn{2}{c}{$\ge 1$ lepton} & \multicolumn{3}{c}{40.9/39.7\%}\\
        \multicolumn{2}{c}{$\ge 6$ jets with $p_T > 30$ GeV} & \multicolumn{3}{c}{37.3/31.8\%}\\
        \multicolumn{2}{c}{$\ge 3$ b-jets with $p_T > 30$ GeV} & \multicolumn{3}{c}{14.3/14.3\%}\\
        \midrule
        && SR-1l-6j-A & SR-1l-6j-B &  SR-1l-6j-C\\
        \multicolumn{2}{c}{$m_T > 140-140-160$ GeV} & 11.3/9.8\% & 11.3/9.8\% & 10.7/9.1\% \\
        \multicolumn{2}{c}{$\slashed{E}_T > 175-225-275$ GeV} & 10.9/9.5\% & 10.0/8.8\% & 8.8/7.4\% \\
        \multicolumn{2}{c}{$\slashed{E}_T/\sqrt{H_T} > 5$ GeV$^{\frac{1}{2}}$} & 10.8/9.4\% & 10.0/8.8\% & 8.8/7.4\% \\
        \multicolumn{2}{c}{$m_{eff}^{incl} > 1000-1000-1500$ GeV} & 10.8/9.4\% & 10.0/8.8\% & 8.8/7.4\% \\
        \bottomrule
    \end{tabular}
    \caption{A comparison of nonzero lepton cut-flow tables between ATLAS and our ``theorist's'' detector for gluinos decaying through a pair of top quarks.} \label{Tab: tt2}
\end{table}

%

\newpage

\bibliographystyle{JHEP}
\renewcommand{\refname}{Bibliography}
\addcontentsline{toc}{section}{Bibliography}
\providecommand{\href}[2]{#2}\begingroup\raggedright

\end{document}